\documentclass[fleqn,twoside,twocolumn,nofootinbib]{revtex4} 
\usepackage[sec]{ujp} 
\begin{document}
\title[TWO-FERMION COMPOSITE QUASI-BOSONS AND DEFORMED OSCILLATORS]
{TWO-FERMION COMPOSITE QUASI-BOSONS\\ AND DEFORMED OSCILLATORS}%
\author{A.M.~Gavrilik}
\affiliation{Bogolyubov Institute for Theoretical Physics, Nat.
Acad. of Sci. of Ukraine}
\address{14b, Metrolohichna Str., Kyiv 03680, Ukraine}
\email{omgavr@bitp.kiev.ua}

\author{I.I.~Kachurik}
\affiliation{Bogolyubov Institute for Theoretical Physics, Nat.
Acad. of Sci. of Ukraine}
\address{14b, Metrolohichna Str., Kyiv 03680, Ukraine}
\email{omgavr@bitp.kiev.ua}

\author{Yu.A.~Mishchenko}
\affiliation{Bogolyubov Institute for Theoretical Physics, Nat.
Acad. of Sci. of Ukraine}
\address{14b, Metrolohichna Str., Kyiv 03680, Ukraine}

\udk{539.1.01;539.12.01}  \razd{\secix}

\newcommand{\be}{\begin{equation}}
\newcommand{\ee}{\end{equation}}
\newcommand{\bea}{\begin{eqnarray}}
\newcommand{\eea}{\end{eqnarray}}
\newcommand {\my}[1]{{\color{blue}#1}}

\setcounter{page}{948}
\maketitle               

\begin{abstract}
The concept of quasi-bosons or composite bosons (like mesons,
excitons, {\it etc}.) has a wide range of potential physical
applications. Even composed of two pure fermions, the quasi-boson
creation and annihilation operators satisfy non-standard commutation
relations. It is natural to try to realize the quasi-boson operators
by the operators of a deformed (nonlinear) oscillator, the latter
constituting a widely studied field of modern quantum physics. In
this paper, it is proven that the deformed oscillators which realize
quasi-boson operators in a consistent way really exist. The
conditions for such realization are derived, and the uniqueness of
the family of deformations under consideration is shown.
\end{abstract}

\section{Introduction}

The study of many-body problems that involve composite particles
essentially differs from those for pointlike particles because of
the necessity of a more complicated treatment. Namely, because of
the internal degrees of freedom due to constituents, the statistical
properties of the composite (thus, not pointlike) particles may
essentially deviate from the purely Bose or Fermi description.
 Such deviation, encapsulated in modified commutation relations,
 can be appropriately modeled (represented) by adopting some deformed
 (say, $q$- or $p,q$-deformed or yet another) version of the oscillator algebra.
 A particular realization of the idea to describe composite bosons
 (or ``quasi-bosons'', see \cite{Per}) in terms of a deformed Heisenberg algebra
 was demonstrated by Avancini and Krein in \cite{Avan}
  who utilized the quonic version \cite{Green} of the deformed boson
  algebra. It should be stressed that these quons differ
  from the widely explored (system of) deformed
  oscillators of the Arik--Coon type \cite{Arik}
  if more than one mode are considered: in that case,
  all modes of the Arik--Coon type are independent
  (that means the mutual commutation of the operators
  corresponding to different modes), unlike the quons
  whose different modes do not commute, see \cite{Green,Avan}.

  Although models of deformed oscillators are known in a diversity of
  versions  \cite{B-M,TD,GR1,p-q,GR4,Jan}, to the best of our
  knowledge,
  a detailed analysis of possible realizations, on their base, of
  quasi-bosons is lacking in the literature.
  The present paper can be considered as a step to fill this gap
  and contains some results in that direction.
  Namely, we carry out the detailed analysis in the important case
  dealing with a set of independent modes (copies) of deformed
  oscillators, whereas, for the individual copy, we examine the most
  general possible structure function $\phi(N)$ of deformation
  which, as is well known, unambiguously determines \cite{Melja,Bona}
  the deformed algebras, i.e. the form of basic commutation relations
  for the annihilation, creation, and number operators, according
  to the formula $ a a^\dagger - a^\dagger a = \phi(N+1) - \phi(N)$.

Diverse models of deformed oscillators, due to their peculiar
properties, have received much attention for the last two decades.
Among the best known and extensively studied deformed oscillator
models, there are such as the $q$-deformed Arik--Coon (AC)
\cite{Arik} and Biedenharn--Macfarlane (BM) ones \cite{B-M}, as well
as the two-parameter $p,\!q$-deformed oscillator \cite{p-q}.
 Besides these, there exists the $q$-deformed Tamm--Dancoff (TD)
 oscillator \cite{TD}, also explored though to a much lesser extent \cite{GR1}.
 Unlike all the mentioned models, there is a very modest knowledge concerning
 the so-called $\mu$-deformed oscillator.
 Introduced in \cite{Jan}, the $\mu$-oscillator shows essentially different
  features and exhibits rather unusual properties \cite{GKR}.

 Being direct extensions of the standard quantum harmonic oscillator,
the deformed oscillators find a diversity of applications in
the description of miscellaneous physical systems involving essential
nonlinearities, from, say, quantum optics and the Landau problem to high-energy
quantum particle phenomenology and modern quantum field
theory (see, e.g., \cite{Man'ko,GR5,AG,gavrS,AGI1,AGI2,Rego1,Rego2}.
 That is why the possible use, in order to realize quasi-bosons,
 of deformed oscillators (deformed bosons) is very desirable due to
a considerable simplification of the relevant analysis, achieved when the
algebra representing the initial system of composite particles reduces
to the algebra corresponding to some deformed oscillator.
 In this sense, the information about the internal structure of particles
 is carried by one or more parameters of deformation.
 The present work realizes the just mentioned procedure of reducing
 for the case of quasi-bosons [2] consisting of two ideal fermions.
 The obtained results are of value since, for {\it composite physical
 particles or quasi-particles} (say, mesons, higgsons, light
even nuclei, excitons, etc.), it would be very useful to have them
realized as deformed bosons, using deformed oscillators.

\section{Quasi-Bosons as Two-Fermion Composites}

We consider, like in~\cite{Avan}, the system of composite boson-like
particles (or quasi-bosons, see~\cite{Per}) such that each copy/mode
of them is composed of two usual fermions.
 First of all, we will study the realization of quasi-bosons
 in terms of a set of independent identical copies of deformed
 AC oscillators~\cite{Arik}.

 By $a^{\dag}_{\mu}$,
$b^{\dag}_{\nu}$, $a_{\mu}$, $b_{\nu},$ we denote, respectively, the creation and
destruction operators of the two (mutually anticommuting) sets of
usual fermions, with standard anticommutation relations,
 and use these fermions to construct quasi-bosons.
 Then, the quasi-bosonic creation and destruction operators
 ${A}^{\dag}_{\alpha},\ {A}_{\alpha}$ (where $\alpha$ labels a particular
 quasi-boson and denotes the whole set of its quantum numbers)
 are given as
\be \label{anzats} {A}^{\dag}_{\alpha}
=\sum\limits_{\mu\nu}\Phi^{\mu\nu}_{\alpha}a^{\dag}_{\mu}b^{\dag}_{\nu},\quad
{A}_{\alpha}
=\sum\limits_{\mu\nu}\overline{\Phi}^{\mu\nu}_{\alpha}b_{\nu}a_{\mu},
\ee
 where $a^{\dag}_{\mu}$, $b^{\dag}_{\nu}$, $a_{\mu}$, $b_{\nu}$
 obey the relations
 \[
 \begin{array}{ll}
 \{a_{\mu},a^{\dag}_{\mu'}\}\equiv
a_{\mu}a^{\dag}_{\mu'}+a^{\dag}_{\mu'}a_{\mu}=\delta_{\mu\mu'},
& \{a_{\mu},a_{\mu'}\}=0,\\[2mm]
\{b_{\nu},b^{\dag}_{\nu'}\}\equiv
b_{\nu}b^{\dag}_{\nu'}+b^{\dag}_{\nu'}b_{\nu}=\delta_{\nu\nu'},
&  \{b_{\nu},b_{\nu'}\}=0 \\
 \end{array}
 \]
 and, in addition, each of $a^{\dag}_{\mu}$, $a_{\mu}$ anticommutes with
 each of $b^{\dag}_{\nu}$,  $b_{\nu}$.
 One can easily check that
\be
 [{A}_{\alpha},{A}_{\beta}]=[{A}^{\dag}_{\alpha},{A}^{\dag}_{\beta}]=0.
\label{2_2} \ee
 For the remaining commutator, we find
 \[
[{A}_{\alpha},{A}^{\dag}_{\beta}]=\delta_{\alpha\beta} -
\Delta_{\alpha\beta},
\]
 where  
\[\nonumber
 \Delta_{\alpha\beta} \equiv
\sum\limits_{\mu\nu\mu'}\overline{\Phi}^{\mu\nu}_{\alpha}
\Phi^{\mu'\nu}_{\beta}a^{\dag}_{\mu'}a_{\mu} +
\sum\limits_{\mu\nu\nu'}\overline{\Phi}^{\mu\nu}_{\alpha}
\Phi^{\mu\nu'}_{\beta}b^{\dag}_{\nu'}b_{\nu} .\]
 For the matrices ${\Phi}_{\alpha},$
 we require the normalization condition
 \be {\rm Tr}({\Phi}_{\alpha} {\Phi}^\dag_\beta)=\delta_{\alpha\beta}.
\label{norm}
 \ee
  The entity $\Delta_{\alpha\beta}$ embodies a deviation
  from the pure bosonic commutation relation.\
  Note that a pure boson (when $\Delta_{\alpha\beta}\!=\!0$) is not
  a particular case of a quasi-boson, because $\Delta_{\alpha\beta}\!=\!0$
  would require $\Phi_{\alpha}\!=\!0$, which would yield the invalidity of
  the very composite structure (\ref{anzats}).

Note that, unlike the realization of quasi-bosonic operators using
the quonic variant of the deformed oscillator algebra, as it was done in
\cite{Avan}, the considered copies of
a deformed oscillator will be completely independent in our treatment below.
 That is, we will assume the validity of (\ref{2_2}) and
$[A_{\alpha},A^{\dag}_{\beta}]=0$ for $\alpha\neq\beta$.

\section{Can the Arik--Coon Type Deformed Oscillators Model the Quasi-Bosons?}

Here, we will model the quasi-bosons by the (independent) system
of $q$-deformed bosons of the Arik--Coon type. The latter obey
 \be
[\mathcal{A}_{\alpha},\mathcal{A}^{\dag}_{\beta}] =
\delta_{\alpha\beta} +
(q^{\delta_{\alpha\beta}}-1)\mathcal{A}^{\dag}_{\beta}\mathcal{A}_{\alpha},
\label{commut} \ee
 where the independence of modes is guaranteed by
$\delta_{\alpha\beta}$.

The quasi-bosonic number operator $\mathcal{N}_{\alpha}$ is defined
as \be \nonumber \mathcal{N}_{\alpha} =
\log_{q}\left(1+(q-1)\mathcal{A}^{\dag}_{\alpha}\mathcal{A}_{\alpha}\right),
\ee which is
 the inversion of  
 $\mathcal{A}^{\dag}\mathcal{A}=\frac{q^\mathcal{N}-1}{q-1}$, see~\cite{Arik}.

We recall that the Arik--Coon model system involves, in addition, the
relations

 These relations imply that the operator
$\mathcal{A}^{\dag}_{\alpha}$ is the raising operator for deformed
bosons
  (correspondingly,  $\mathcal{A}_{\alpha}$ -- lowering operator).

Our goal is to find such coefficients $\Phi^{\mu\nu}_{\alpha}$  that
realization  (\ref{anzats}) is in agreement with (\ref{commut}),
   i.e. that relation  (\ref{commut})
   is valid on the appropriate space of states.
 If the ground state $|O\rangle$ for quasi-bosons is defined as
 \be\nonumber
A_{\alpha}|O\rangle=a_{\mu}|O\rangle=b_{\nu}|O\rangle=0,
 \ee
 then the respective space of states is nothing but the
linear span
$\{|O\rangle,A^{\dag}_{\gamma_1}|O\rangle,A^{\dag}_{\gamma_2}A^{\dag}_{\gamma_1}
|O\rangle,\ldots\}$.
 Rewrite the commutation relations (\ref{commut}) as
  \be \nonumber
  F_{\alpha\beta}\equiv\Delta_{\alpha\beta}
+ (q^{\delta_{\alpha\beta}}-1)A^{\dag}_{\beta}A_{\alpha} =0. \ee
   Then the validity of commutation relations on the indicated
 linear span reduces to nullifying
 each of the states $|O\rangle$, $A^{\dag}_{\gamma_1}|O\rangle$,
$A^{\dag}_{\gamma_2}A^{\dag}_{\gamma_1}|O\rangle$, ... by the operator $F_{\alpha\beta}$.

Obviously, for the ground state, we have
  \be\nonumber
F_{\alpha\beta}|O\rangle=0.
 \ee
 Notice that
\begin{align}\nonumber
&F_{\alpha\beta}A^{\dag}_{\gamma_1}|O\rangle=0  &
&\Leftrightarrow\quad
[F_{\alpha\beta},A^{\dag}_{\gamma_1}]|O\rangle=0\, ,\\ \nonumber
&F_{\alpha\beta}A^{\dag}_{\gamma_1}A^{\dag}_{\gamma_2}|O\rangle=0
 & &\Leftrightarrow\quad
 [[F_{\alpha\beta},A^{\dag}_{\gamma_1}],A^{\dag}_{\gamma_2}]|O\rangle=0.
\end{align}
The equality $[F_{\alpha\beta},A^{\dag}_{\gamma_1}]|O\rangle=0$ can
be rewritten in the form of a relation on matrices $\Phi_{\alpha}$.
Using this, the commutator reduces to
 \be \nonumber
[F_{\alpha\beta},A^{\dag}_{\gamma_1}] =
(1-q^{\delta_{\alpha\beta}})A^{\dag}_{\beta}\left[F_{\alpha\gamma_1}
+ (1-q^{\delta_{\alpha\gamma_1}})
A^{\dag}_{\gamma_1}A_{\alpha}\right]. \ee
 Calculate the double commutator:
\begin{align*}
&[[F_{\alpha\beta},A^{\dag}_{\gamma_1}],A^{\dag}_{\gamma_2}] =
(1-q^{\delta_{\alpha\beta}})A_{\beta}^{\dag}
[F_{\alpha\gamma_1},A^{\dag}_{\gamma_2}] +\\
&+(1-q^{\delta_{\alpha\beta}})(1-q^{\delta_{\alpha\gamma_1}})
A_{\beta}^{\dag}A_{\gamma_1}^{\dag}[A_{\alpha},A^{\dag}_{\gamma_2}].
\end{align*}
 From this equation, for the relation
$[[F_{\alpha\beta},A^{\dag}_{\gamma_1}],A^{\dag}_{\gamma_2}]|O\rangle=0$
 to hold, we infer
  \be\nonumber
(1-q^{\delta_{\alpha\beta}})(1-q^{\delta_{\alpha\gamma_1}})\delta_{\alpha\gamma_2}
A_{\beta}^{\dag}A_{\gamma_1}^{\dag}|O\rangle=0.
 \ee
 Thus, we come to the contradiction: at
$\alpha=\beta=\gamma_1=\gamma_2$ and $q\neq 1,$ it follows that
 \be\nonumber
(A_{\alpha}^{\dag})^2|O\rangle=0, \ee
 the paradoxical fact -- nilpotency of ``bosonic''  operators.

    Hence, the Arik--Coon type deformation, see (\ref{commut}),
 leads to the inconsistency and so is inappropriate for a realization of
 quasi-bosons.
   The situation changes, however, for other deformations,
 as will be seen below.

\section{Quasi-Bosons vs Deformation of General Form}

 In what follows, 
we study the independent quasi-bosons system realized by deformed
oscillators {\it without indication of a particular model of
deformation}.
 In this section, we obtain the necessary conditions for such realization
 in terms of the structure function and matrices $\Phi_{\alpha}$.

Let $\phi$ be the structure function of deformation.
 The quasi-boson number operator is introduced as 
\be\nonumber N_{\alpha}\mathop{=}\limits^{\rm def}
\phi^{-1}(A^{\dag}_{\alpha}A_{\alpha}). \ee
  Note that this definition is not unique.
 Another equivalent definition could be given, e.g.,
  $N_{\alpha}\mathop{=}\limits^{\rm def}\phi^{-1}(A_{\alpha}A^{\dag}_{\alpha})-1.$
  Below,
  we need the notion of weak equality denoted by the symbol $\cong$.
Namely, if $G$ is some operator function, then its weak equality (to
zero) means
 \be
G(A,A^{\dag},N;...)\cong 0 \ \mathop{\Leftrightarrow}\limits^{\rm
def} G(...) A^{\dag}_{\gamma_m}...\, A^{\dag}_{\gamma_1}|O\rangle
=0\ \label{slabrav}\ee
 for $m=0,1,2, ...\, .$

\subsection{Derivation of necessary conditions}

 We require the validity of the following weak equalities for commutators:
\begin{align}\nonumber
&[N_{\alpha},A^{\dag}_{\alpha}]\cong
    A^{\dag}_{\alpha},\quad [N_{\alpha},A_{\alpha}]\cong -A_{\alpha}
,\\
&[A_{\alpha},A^{\dag}_{\beta}]\cong 0 \quad\text{if } \alpha\neq\beta,\label{system1}\\
&[A_{\alpha},A^{\dag}_{\alpha}] \cong
\phi(N_{\alpha}+1)-\phi(N_{\alpha}).\nonumber
\end{align}
  We also emphasize that, whatever is the definition of $N_{\alpha}$,
  the following implications must be true:
\begin{align*}
&\phi(N_{\alpha})\cong A^{\dag}_{\alpha}A_{\alpha} &\Rightarrow&\quad\phi(0)=0,
                \qquad\qquad\\
&\phi(N_{\alpha}+1)\cong A_{\alpha}A^{\dag}_{\alpha} &\Rightarrow&\quad\phi(1)=1.
                 \qquad\qquad
\end{align*}
 From the second relation in (\ref{system1}),  the equality
\be
\sum\limits_{\mu'\nu'}\left(\Phi^{\mu\nu'}_{\beta}\overline{\Phi}^{\mu'\nu'}_{\alpha}\Phi^{\mu'\nu}_{\gamma}
+
\Phi^{\mu\nu'}_{\gamma}\overline{\Phi}^{\mu'\nu'}_{\alpha}\Phi^{\mu'\nu}_{\beta}\right)
=0,\ \ \alpha\neq\beta,\label{uslnez} \ee
 does follow, which can be rewritten in the matrix form
 \be
\Phi_{\beta}\Phi^{\dag}_{\alpha}\Phi_{\gamma}+
\Phi_{\gamma}\Phi^{\dag}_{\alpha}\Phi_{\beta}=0,\quad
\alpha\neq\beta.\label{req1} \ee
 Since
$A^{\dag}_{\alpha}A_{\alpha}\cong\phi(N_{\alpha})$ and
$A_{\alpha}A^{\dag}_{\alpha}\cong\phi(N_{\alpha}+1)$,
 we have
 \be  \nonumber
 [A^{\dag}_{\alpha}A_{\alpha},A_{\alpha}A^{\dag}_{\alpha}]\cong
 0 \qquad {\rm and} \qquad
 [\Delta_{\alpha\alpha},N_{\alpha}]\cong 0.
 \ee
 The first of these relations can be   
 rewritten as
  \[
[A^{\dag}_{\alpha}A_{\alpha},\Delta_{\alpha\alpha}]\cong
0.      
 \]

After calculations, this commutator takes the form
\begin{align}\nonumber
&[A^{\dag}_{\alpha}A_{\alpha},\Delta_{\alpha\alpha}]=
2A^{\dag}_{\alpha}\sum\limits_{\mu\nu}\left(\Phi_{\alpha}\Phi^{\dag}_{\alpha}
\Phi_{\alpha}\right)^{\dag}_{\nu\mu}b_{\nu}a_{\mu}-\\
&-2\sum\limits_{\mu'\nu'}\left(\Phi_{\alpha}
\Phi^{\dag}_{\alpha}\Phi_{\alpha}\right)_{\mu'\nu'}a^{\dag}_{\mu'}b^{\dag}_{\nu'}
A_{\alpha}\cong 0.\label{w_e}
\end{align}
 We denote the matrix in parentheses by $\Psi_{\alpha}\equiv
\Phi_{\alpha}\Phi^{\dag}_{\alpha}\Phi_{\alpha}$.
 For the weak equality in (\ref{w_e}) to be valid, it is
 necessary that the following commutator with the creation operator
 give zero on the vacuum state:
\[
\left[(\overline{\Psi}^{\mu\nu}_{\alpha}\Phi^{\mu'\nu'}_{\alpha} -
\overline{\Phi}^{\mu\nu}_{\alpha}\Psi^{\mu'\nu'}_{\alpha})
a^{\dag}_{\mu'}b^{\dag}_{\nu'}b_{\nu}a_{\mu},\Phi_{\alpha}^{\lambda\rho}
a^{\dag}_{\lambda}b^{\dag}_{\rho}\right]|O\rangle=\]
    \vspace{-7mm}
 \be =\left(\Phi_{\alpha}^{\mu'\nu'}\cdot
\text{Tr}(\Psi_{\alpha}^{\dag}\Phi_{\alpha}) -
\Psi_{\alpha}^{\mu'\nu'}\right)a^{\dag}_{\mu'}b^{\dag}_{\nu'}|O\rangle=0.
\ee
 This leads to the requirement
 \be
\Phi_{\alpha}\Phi^{\dag}_{\alpha}\Phi_{\alpha} =
\text{Tr}(\Phi^{\dag}_{\alpha}\Phi_{\alpha}\Phi^{\dag}_{\alpha}\Phi_{\alpha})\cdot
\Phi_{\alpha},\label{req2} \ee
 which is also the sufficient one.
  Then we come to two  requirements, (\ref{req1}) and (\ref{req2}),
for the matrices $\Phi_{\alpha}.$

\subsection{Relating \boldmath$\Phi_{\alpha}$ to the structure function
\boldmath$\phi(n)$}

Now let us derive the relations that involve the structure function
$\phi$. For the commutator $[A_{\alpha},A^{\dag}_{\alpha}],$ we have
\[
[A_{\alpha},A^{\dag}_{\alpha}] = 1-\Delta_{\alpha\alpha} \cong
\phi(N_{\alpha}+1)-\phi(N_{\alpha}).
\]
  From the latter,
 \be\nonumber F_{\alpha\alpha} \equiv
\Delta_{\alpha\alpha} - 1 + \phi(N_{\alpha}+1)-\phi(N_{\alpha})
\cong 0. \ee
 If the conditions
(see the first line in (\ref{system1}))
\be [N_{\alpha},A^{\dag}_{\alpha}]\cong A^{\dag}_{\alpha},\quad
[N_{\alpha},A_{\alpha}]\cong -A_{\alpha}\label{n_commut} \ee do hold
(this means that, for these conditions, a verification is needed,
see Sec. 4.3 below),
 then \[
\phi(N_{\alpha})A^{\dag}_{\alpha}\cong A^{\dag}_{\alpha}
\phi(N_{\alpha}+1)\quad\Rightarrow \]
 \vspace{-7mm}
 \[\quad \Rightarrow\
[\phi(N_{\alpha}),A^{\dag}_{\alpha}]\cong A^{\dag}_{\alpha}
\bigl(\phi(N_{\alpha}+1)-\phi(N_{\alpha})\bigr). \]
 That leads to the relation
 \[
[F_{\alpha\alpha},A^{\dag}_{\alpha}] \cong
2(\Phi_{\alpha}\Phi^{\dag}_{\alpha}\Phi_{\alpha})^{\mu\nu}a^{\dag}_{\mu}b^{\dag}_{\nu}
+\]
  \vspace{-7mm}
 \be      \label{5_5}
 + A^{\dag}_{\alpha} \Bigl(\phi(N_{\alpha}+2) -
2\phi(N_{\alpha}+1) + \phi(N_{\alpha})\Bigr). \ee
  From the requirement that this commutator vanish on the vacuum state,
   we obtain (note that $\phi(0)=0$):
  \be\nonumber
\Phi_{\alpha}\Phi^{\dag}_{\alpha}\Phi_{\alpha} =
\Bigl(\phi(1)-\frac12 \phi(2)\Bigr)\Phi_{\alpha} = \frac{f}{2}
\Phi_{\alpha}, \ee where the (deformation) parameter $f$ is
introduced:
\[
\frac{f}{2}\equiv \phi(1)-\frac12 \phi(2)=
{\mathrm{Tr}}(\Phi^{\dag}_{\alpha}\Phi_{\alpha}\Phi^{\dag}_{\alpha}\Phi_{\alpha})
 \ \ {\mathrm{for~all}} \ \    \alpha.
\]
 Then equality (\ref{5_5}) takes the form
\[
[F_{\alpha\alpha},A^{\dag}_{\alpha}] \cong
f\cdot A^{\dag}_{\alpha} + A^{\dag}_{\alpha} \bigl(\phi(N_{\alpha}+2) - 2\phi(N_{\alpha}+1) +
\phi(N_{\alpha})\bigr).
\]
  By induction, the equality for the
$n$-th commutator ($C_n^k$ denote  binomial coefficients) can be
proven:
  \be\nonumber [...[F_{\alpha\alpha},A^{\dag}_{\alpha}]...
A^{\dag}_{\alpha}] \!\cong \!(A^{\dag}_{\alpha})^n
\biggl\{\sum\limits_{k=0}^{n+1}\!(\!-\!1)^{n\!+\!1\!-\!k}
C^k_{n+1}\phi(N_{\alpha}\!+k) \biggr\}. \ee
 Using the requirement that the $n$-th commutator
vanish on the vacuum state, we derive the recurrence relation
  \be
\phi(n+1) = \sum\limits_{k=0}^{n}(-1)^{n-k} C^k_{n+1}\phi(k) , \quad
n\geq 2.\label{recurr1} \ee
  Thus, all the values $\phi(n)$ for $n\geq 3$ are determined
  unambiguously by the values $\phi(1)$ and $\phi(2)$ which depend,
  in general, on some set of deformation parameters.

It can easily be shown that the non-deformed structure function
 $\phi(n)\equiv n$ satisfies Eq. (\ref{recurr1}).
  Similarly, one proves the following  natural ``initial'' conditions:
\be \nonumber \phi(1)\rightarrow 1,\quad \phi(2)\rightarrow
2\quad\Rightarrow\quad \forall k>2\quad \phi(k)\rightarrow k,
\ee
when all the deformation parameters tend to their non-deformed
values.

Taking into account the equality \cite{Korn}
 \be\nonumber
\sum\limits_{k=0}^{n}C_n^k k^m (-1)^{n-k} =
\begin{cases}
0,\quad m<n,\cr n!, \ \ m=n,
\end{cases}
\ee
 we see that the only independent solutions
 of the recurrence relation (\ref{recurr1}) are $n$ and $n^2$,
 as well as their linear combination
\be \phi(n)=\left(1+\frac{f}{2}\right)n -
\frac{f}{2}n^2.\label{solution1}
  \ee
  This formula satisfies both the initial conditions and
  the recurrence relations (\ref{recurr1}).
  In view of the uniqueness of the solution under
fixed initial conditions, formula (\ref{solution1}) gives
the general solution of relation (\ref{recurr1}).

\subsection{Verification of relations (\ref{n_commut})}

\vspace{1mm} As we have mentioned above, it remains to satisfy
relations (\ref{n_commut}).
  Note that the second of them stems by conjugation from the
first one,
\be \label{usl2} [N_{\alpha},A^{\dag}_{\alpha}]\cong
A^{\dag}_{\alpha}. \ee
  In view of the {\em independence} of different modes, see
 (\ref{uslnez}), it is enough to set $\gamma_1=\gamma_2=\ldots=\alpha$
 on the r.h.s. of (\ref{slabrav}).

Let us denote, by $I_n,$ the operators
\begin{gather}
I_0\!=\!N\!\equiv\! \phi^{-1}(A^{\dag}_{\alpha}A_{\alpha}),\nonumber\\
I_{n+1}=[I_n,A^{\dag}_{\alpha}]=[...[N_{\alpha},A^{\dag}_{\alpha}]...
A^{\dag}_{\alpha}]. \label{5_22}
\end{gather}
In terms of these operators, Eq. (\ref{usl2}) is written as
\be I_1|O\rangle=A^{\dag}_{\alpha}|O\rangle,\quad
I_n|O\rangle=0,\quad n>1.\label{q3} \ee
 Introduce the notation
\be\nonumber \varepsilon_{\alpha}\equiv 1-\Delta_{\alpha\alpha}=
[A_{\alpha},A^{\dag}_{\alpha}]. \ee
 Using the auxiliary relations
\[ [\Delta_{\alpha\alpha},A^{\dag}_{\alpha}]=f
A^{\dag}_{\alpha},\quad \ [\Delta_{\alpha\alpha},A_{\alpha}]=
-\overline{f}A_{\alpha},\]
   \vspace{-7mm}
\[
[\varepsilon_{\alpha},A^{\dag}_{\alpha}]=-fA^{\dag}_{\alpha},\quad
[\Delta_{\alpha\alpha},N_{\alpha}]\cong 0,\quad
\Delta_{\alpha\alpha}=\Delta_{\alpha\alpha}^{\dag},
 \]
  we come to the equalities
  \be
\left[(A^{\dag}_{\alpha}A)^n,A^{\dag}_{\alpha}\right]=
A^{\dag}_{\alpha}\left[(A^{\dag}_{\alpha}A_{\alpha}+
\varepsilon_{\alpha})^n-(A^{\dag}_{\alpha}A_{\alpha})^n\right],\label{e1}\ee
  \vspace{-7mm}
\be
\left[\varepsilon_{\alpha}^n,A^{\dag}_{\alpha}\right]=A^{\dag}_{\alpha}
[(-f+\varepsilon_{\alpha})^n-\varepsilon_{\alpha}^n]. \label{e2} \ee
 From these equalities, we derive the appropriate
expression for the $n$-fold commutator (\ref{5_22})
($\alpha_n~=~n(n-1)/2$):
 \[
I_n=(A^{\dag}_{\alpha})^n\phi^{-1}(A^{\dag}_{\alpha}A_{\alpha}+
n\varepsilon_{\alpha}-\alpha_n
f) - \sum\limits_{k=0}^{n-1}C_n^k (A^{\dag}_{\alpha})^{n-k}I_k .
 \]
  Finally, conditions (\ref{q3}) can be cast in the form
 \[
A^{\dag}_{\alpha}\phi^{-1}
(A^{\dag}_{\alpha}A_{\alpha}+\varepsilon_{\alpha})|O\rangle=A^{\dag}_{\alpha}|O\rangle,
 \] 
\vspace{-7mm}
 \[
(A^{\dag}_{\alpha})^n\phi^{-1}(A^{\dag}_{\alpha}A+n\varepsilon_{\alpha}-\alpha_n
f)|O\rangle = n(A^{\dag}_{\alpha})^{n}|O\rangle,\  n>1.
\]
  To satisfy the first of these equalities, we require that
  \be\nonumber
\phi^{-1}(1)=1\quad\Rightarrow\quad \phi(1)=1. \ee
  Likewise, the second equality to be valid requires:\
  $ \phi^{-1}\Bigl(n\!-\!\frac{n(n-1)}{2}f\Bigr)\!=\!n$.
 This gives us the ``bonus'' in the form of expression (\ref{solution1})
 for the structure function.

 {\em Remark.} Using the obtained results, it is easy to
  derive the recurrence relation for the structure function,
\[
\phi(n+1)=\frac{2(n+1)}{n}\phi(n)-\frac{n+1}{n-1}\phi(n-1),\]
 and, for the Hamiltonian $H=\frac12(\phi(n+1)+\phi(n))$,
 the recurrence relation for its eigenvalues (the energies):
\[
E_{n+1}=\frac{4n^2+4n-4}{2n^2-1}E_{n}-\frac{2n^2+4n+1}{2n^2-1}E_{n-1}.
\]
  The latter  has typical form of the so-called
quasi-Fibonacci relation \cite{GKR}. The general class of deformed
oscillators with polynomial structure functions $\phi(N)$  
 (these are quasi-Fibonacci as well) was studied in
\cite{GR5}.

\section{Admissible Matrices \boldmath$\Phi_{\alpha}$ }

It remains to find the admissible matrices $\Phi_{\alpha}$.
 These should satisfy (\ref{norm}) and the equations
 \bea
 \label{sys1}
 \hspace{-0.5cm}&&\Phi_{\alpha}\Phi^{\dag}_{\alpha}\Phi_{\alpha}=(f/2)
              \Phi_{\alpha},\\
 \hspace{-0.5cm}&&\Phi_{\beta}\Phi^{\dag}_{\alpha}\Phi_{\gamma}+
\Phi_{\gamma}\Phi^{\dag}_{\alpha}\Phi_{\beta}=0,
           \alpha\neq\beta. \label{sys2}
 \eea
 Let us assume $f\ne 0$.
  If $\det \Phi_{\alpha}\neq 0$ for some-$\alpha$,
 Eq. (\ref{sys1}) yields
 \be \nonumber\Phi_{\alpha}\Phi^{\dag}_{\alpha}=\frac{f}{2}{\bf 1} . 
  \ee
   From Eq. (\ref{sys2}) at $\gamma=\alpha,$ we  obtain
  \be\nonumber
  \Phi_{\beta}=0,\quad\forall\beta\neq\alpha.
  \ee
   Then it follows that only one value of $\alpha$ is possible, for which $\det
\Phi_{\alpha}\neq 0$.
 In that case, $\Phi_{\alpha}$ is an arbitrary unitary matrix.
 All the rest $\Phi_{\beta}=0,\ \beta\neq\alpha$.
That gives the partial non-degenerate solution of the system.
 All other solutions will be degenerate for all $\alpha$.

 Now what concerns the case of degenerate solutions.
  Using some facts from linear algebra (the Fredholm theorem, {\it etc}.),
  we come to the following implication:
 \[
{\mathrm{Tr}}(\Phi_{\alpha}\Phi^{\dag}_{\alpha})=1 \ \ \Rightarrow\
\ {\mathrm{rank}}\,(\Phi_{\alpha})=2/f\equiv m\ {\mathrm{for~all}} \
\alpha.
\]
 So, the deformation parameter $f$ has a discrete range
of values (if the set of indices $\mu,\nu$ is finite or enumerable):
\vspace{-4mm}
 \be      
  f=\frac2m  \ \ \ \Rightarrow \ \ \
     \phi(n)=\Bigl(1+\frac1m\Bigr)n-\frac1m n^2.
\vspace{-4mm}
 \ee
 The set of the solutions depends on the relation
between $d$ and $k\!\cdot\! m$, where $k$ denotes the number of
independent copies (modes) of deformed bosons, and $d$ is the
smallest dimension among the dimensions of matrices $\Phi_{\alpha}$.
 If \mbox{$d\!\cdot\! m\!>\!d$}, the set of solutions is empty.
  If $k\!\cdot\!m\leq d$, then there exist such unitary
 matrices $U_1$ and $U_2$ that the following matrix
product is block-diagonal:
  \vspace{-1mm}
  \be\nonumber
U^{\dag}_1\Phi_{\alpha}U_2=
\begin{pmatrix}
0 & 0 & 0\\
0 & \tilde{\Phi}_{\alpha} & 0\\
0 & 0 & 0
\end{pmatrix}.
\ee
 Then the $m\times m$ matrix $\tilde{\Phi}_{\alpha}$  obeys the
equation
 \be\nonumber
\tilde{\Phi}_{\alpha}\tilde{\Phi}^{\dag}_{\alpha}=
  \frac{f}{2}{\bf 1}.
 \ee
  Its general solution can be given through the unitary
matrix
\be\nonumber \tilde{\Phi}_{\alpha}= \sqrt{f/2}\ 
U_{\alpha}(m). \ee
 Thus, the general solution of Eqs. (\ref{sys1}) and (\ref{sys2})
 is   
 \be
 \Phi_{\alpha}=U_1
\text{diag}\biggl\{0,\sqrt{\frac{f}{2}}U_{\alpha}(m),0\biggr\}U^{\dag}_2,
\label{gen_solution} \ee
 where, for every matrix $\Phi_{\alpha}$, the block
$\sqrt{\frac{f}{2}}U_{\alpha}(m)$ in (\ref{gen_solution}) is at
the $\alpha$-th place and has zero intersection with the corresponding
block of any other matrix $\Phi_{\beta}$ for $\beta\neq \alpha$.

\section{Concluding Remarks}

 Here, we make a kind of resume, also pointing out some further
directions. For the system of completely independent quasi-bosons,
their representation in terms of deformed bosons of the AC type
fails. Nevertheless, the desired realization is possible with some
other structure function $\phi$ of the form (\ref{solution1}),
i.e. with the structure function which is {\it quadratic in the
number operator} and contains one parameter of deformation. The
additional very important necessary and sufficient conditions on
the matrices $\Phi_{\alpha}$ involved in construction
(\ref{anzats}) of quasi-bosons, for such representation to be
consistent, are derived. They can be completely solved which
results in the general solution (\ref{gen_solution}).

Although we used pure fermions as constituents, the analysis shows
that the parameter of deformation giving the quasi-boson's
realization (see (23)) is linked with a discrete characteristic $m$
(the rank) of the matrix $\Phi_\alpha$.

 As the further nearest goals, it
is interesting to study more complicated situations.
  First of all, it is natural to extend the
construction of quasi-bosons, formed from two particles, to the case
of the constituents that are not fermions but a (particular or
general) deformation of fermions. Some results already obtained in
this direction will be published separately. Another path of the
extension is the treatment of quasi-independent quasi-bosons, in
which case one should start with a proper definition of the
``physical'' subspace of quasi-bosonic states.\looseness=1

{\it Note added in proof.} Recently, the above results have been
extended to the case of quasi-bosons composed of two $q$-deformed
fermions \cite{q-ferm}. In addition, using the results of the
present work, the relation between the entanglement in composite
(fermion+fermion) bosons realized by deformed bosons and the
parameter of deformation, is established \cite{Ent}.

\vskip3mm
 This research was partially supported by Grant 29.1/028 of
 the State Foundation of Fundamental Research of Ukraine
and by the Special Program of the Division of Physics and Astronomy
of the NAS of Ukraine.

\rezume{КВАЗІБОЗОНИ, СКЛАДЕНІ З ДВОХ ФЕРМІОНІВ,\\ ТА ДЕФОРМОВАНІ
ОСЦИЛЯТОРИ} {О.М. Гаврилик, І.І. Качурик, Ю.А. Міщенко} {Поняття
квазібозонів чи складених бозонів має широкий спектр
 фізичних застосувань (мезони, ексітони тощо). Відомо, що навіть
 у випадку квазібозонів, складених із двох звичайних фермі\-онів,
 їх оператори народження і знищення задовольняють нестандартні
 комутацій\-ні співвідношення.   Природно спробува\-ти реалізувати
 квазібозонні операто\-ри відповідно операторами народження і знищення
 деформова\-них (нелінійних) осциля\-торів, адже останні становлять
 добре вивчену область сучасної квантової фізики.
 У статті доведено, що такі деформовані осцилятори, які реалізують
квазібозони, справді існують. Виве\-дено необхідні і достатні умови
для 
реалізації. Також доведено єдиність сім'ї можливих деформацій.}

\end{document}